\documentclass{ieeeaccess}
\usepackage{cite}
\usepackage{amsmath,amssymb,amsfonts}
\usepackage{algorithmic}
\usepackage{graphicx}
\usepackage{textcomp}
\usepackage{caption}
\usepackage{array}
\usepackage{enumitem}
\usepackage{tabularx,booktabs}
\usepackage{tabularray}
\usepackage{comment}
\def\BibTeX{{\rm B\kern-.05em{\sc i\kern-.025em b}\kern-.08em
    T\kern-.1667em\lower.7ex\hbox{E}\kern-.125emX}}
\begin{document}
\history{Date of publication xxxx 00, 0000, date of current version xxxx 00, 0000.}
\doi{10.1109/ACCESS.2017.DOI}

\newcolumntype{C}{>{\centering\arraybackslash}X} 
\setlength{\extrarowheight}{1pt} 

\title{A Comparative Performance Analysis of Explainable Machine Learning Models With And Without RFECV Feature Selection Technique Towards Ransomware Classification}
\author{\uppercase{Rawshan Ara Mowri}\authorrefmark{1},
\uppercase{Madhuri Siddula\authorrefmark{2}, and Kaushik Roy}\authorrefmark{3}.}
\address[1,2,3]{Department of Computer Science, North Carolina A\&T State University, Greensboro, NC 27411 USA}
\tfootnote{This work is funded by NetApp.}

\markboth
{Author \headeretal: Preparation of Papers for IEEE TRANSACTIONS and JOURNALS}
{Author \headeretal: Preparation of Papers for IEEE TRANSACTIONS and JOURNALS}

\corresp{Corresponding author: Rawshan Ara Mowri (e-mail: rmowri@aggies.ncat.edu).}

\begin{abstract}
Ransomware has emerged as one of the major global threats in recent days. The alarming increasing rate of ransomware attacks and new ransomware variants intrigue the researchers in this domain to constantly examine the distinguishing traits of ransomware and refine their detection or classification strategies. Among the broad range of different behavioral characteristics, the trait of Application Programming Interface (API) calls and network behaviors have been widely utilized as differentiating factors for ransomware detection, or classification. Although many of the prior approaches have shown promising results in detecting and classifying ransomware families utilizing these features without applying any feature selection techniques, feature selection, however, is one of the potential steps toward an efficient detection or classification Machine Learning model because it reduces the probability of overfitting by removing redundant data, improves the model’s accuracy by eliminating irrelevant features, and therefore reduces training time. There have been a good number of feature selection techniques to date that are being used in different security scenarios to optimize the performance of the Machine Learning models. Hence, the aim of this study is to present the comparative performance analysis of widely utilized Supervised Machine Learning models with and without RFECV feature selection technique towards ransomware classification utilizing the API call and network traffic features. Thereby, this study provides insight into the efficiency of the RFECV feature selection technique in the case of ransomware classification which can be used by peers as a reference for future work in choosing the feature selection technique in this domain.
\end{abstract}


\titlepgskip=-15pt

\maketitle

\section{Introduction}
\label{sec:introduction}
\PARstart{R}{ansomware} is a harmful software that applies symmetric and asymmetric cryptography to inscribe user information and poses a Denial-of-Service (DoS) attack on the intended user [1]. The unique functional process of ransomware attacks makes it more harmful than any malware attacks and causes irreversible losses. Crypto-viral Extortion’, which is the functional process of ransomware, includes three main steps [2] as depicted in Figure 1. In the initial step, the attacker creates a key pair that incorporates a private key K1 and a public key K2, puts the public key K2 in the ransomware, then, at that point, launches the ransomware. After entering a computer, in the second step, the ransomware activates itself and produces an arbitrary symmetric session key K3 to encrypt the victim’s files or data. Next, the ransomware utilizes K2 to encrypt K3 and to create a small irregular ciphertext E1. Then, the ransomware zeroizes K3 and the plaintext from the person's drive. A communication bundle P1 containing previously generated E1, a payment note M, and a medium to contact the attacker, is then created. After that, the ransomware informs the victim of the attack and demands payment via a transaction medium within a set amount of time in order to decrypt the files by displaying the payment note M. At the final step, as the payment is completed, the communication bundle P1 is adjusted to P2 containing just the deviated ciphertext E1 and steered back to the attacker. The attacker gets P2, decrypts E1 with K1, and gets K3 which is then sent back to the victim to decrypt the files. Finally, upon receiving K3, the victim decrypts the files. Usually, the victim pays the ransom using untraceable cryptocurrency [3]. However, paying the ransom doesn’t guarantee that the decryption key could secure the encrypted files, which could be the worst scenario of any type of ransomware attack [4].

Supported by a report by Symantec in 2015, there are two types of ransomware [5]-
\begin{itemize}
    \item Locker ransomware: denies access to the system or device
    \item Crypto ransomware: denies access to the files or data
\end{itemize}

However, according to [6], based on the functionalities, ransomware is categorized into four groups-

\begin{itemize}
    \item Encrypting ransomware: encrypts and denies access to the victim’s files and data (i.e., AIDS Trojan, CryptoLocker, WannaCry, CryptoWall) [6]
    \item Non-encrypting ransomware: doesn’t do encryption but rather threatens to try if the ransom is not paid (i.e., WinLock, NotPetya) [6]
    \item Leak-ware: doesn’t do encryption instead claims to reveal stolen information from the victim’s system if the ransom is not paid [7]
    \item Mobile ransomware: targets the Android platform [8]
    
\end{itemize}

\begin{figure}[htb]
	\makebox[\linewidth][c]{\includegraphics[angle = 0, clip, trim=0cm 0cm 0cm 0cm, width=0.5\textwidth]{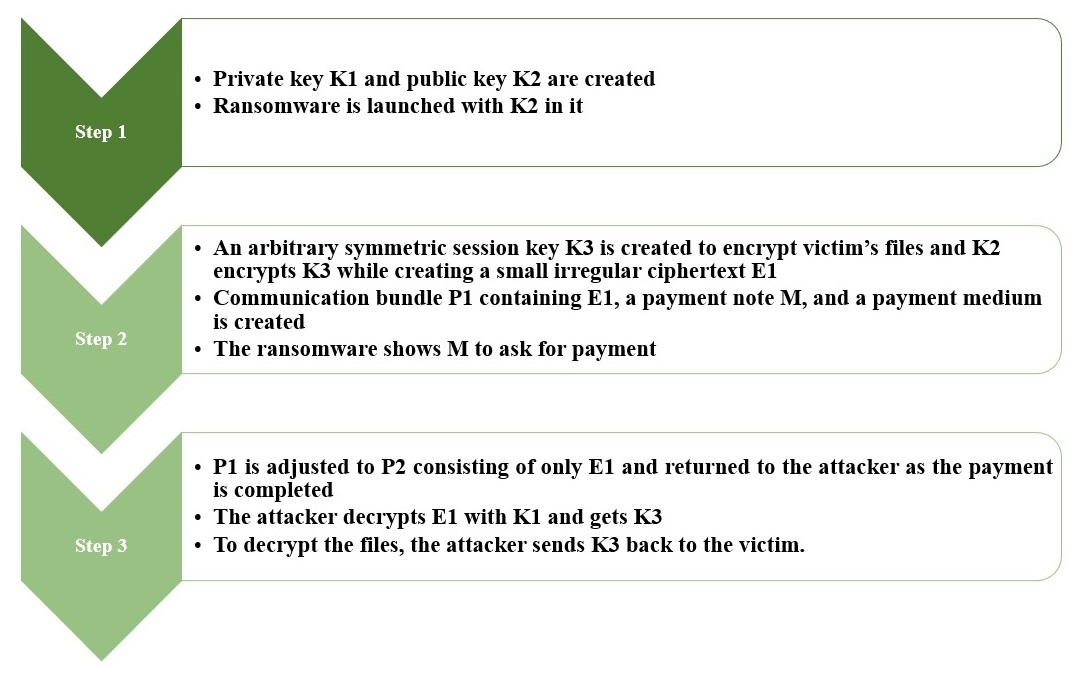}}
        \caption{Workflow of a ransomware}
	\label{fig:pic1}
	\vspace{-10pt}
\end{figure}

All these categories of ransomware are playing a vital role in the recent upsurge in the incidence of ransomware attacks. Due to the increasing number of ransomware variants and ransomware attacks, researchers have been earnestly involving themselves to look for efficient ways to improve the scenarios. While some researchers are analyzing the distinctive behaviors of ransomware by executing it in a secure environment called Dynamic Analysis [1], [9] - [13], some researchers are analyzing the ransomware without any execution, referred to as Static Analysis [14] - [16]. However, a good number of researchers are combining these two approaches and adopting a Hybrid Analysis Approach [17] - [19]. Although the static analysis technique takes less analysis time and facilitates the researchers by not requiring the execution of malicious files, this technique struggles to trace new ransomware variants because of the ever-evolving code obfuscation technique. On the other hand, although a dynamic analysis approach might take a longer time to process and analyze the ransomware program, this approach can detect ransomware with higher accuracy as it executes the ransomware program in a secure virtual environment and does real-time behavioral analysis. The main idea is that despite the changes in the new ransomware variants, they will still show the same behavioral patterns. Therefore, for this study, we have opted for the dynamic analysis approach for its ability to detect and classify ransomware families based on behavioral patterns regardless of the code obfuscation techniques deployed by the ransomware programmers [20], [21].


The main contributions of this study are:
\begin{itemize}
    \item Developing a Web-Crawler, ‘GetRansomware’ to automate collecting the Windows Portable Executable (PE) files of 15 different ransomware families from the VirusShare repository. The Web-Crawler is essential to automate searching and downloading the samples and cutting down the manual workload. 
    \item Examining and comparing the performance of six Supervised Machine Learning models with and without RFECV feature selection technique in case of classifying ransomware families. For this task, we construct two different datasets by analyzing two types of binaries, namely, Windows Portable Executables (PE) and Packet Capture (PCAP) files. Since our approach includes utilizing RFECV for selecting the optimum number of features and RandomSearchCV for selecting the optimum hyperparameter values for each classifier, therefore, this study attempts to optimize each model's performance in both scenarios before the comparison is made. 
    \item Presenting the efficiency of the RFECV feature selection technique in ransomware classification with respect to the performance of the Machine Learning models. For this
    task, first, we utilize ‘Shapley Additive exPlanations’ to obtain the highly contributing features
    from the without feature selection scenario. Next, we obtain the RFECV-selected features from the with feature selection scenario. Finally, we report how the important set of features varies for each Machine Learning model in two scenarios and how they affect to the final outcome.
    
\end{itemize}

The rest of this paper is structured as follows: Section 2 presents the related works. Section 3 details our methodology. The experimental results and discussions are illustrated in Section 4. Section 5 concludes the paper with the direction for future works.

\section{Related Works}
In this section, we present several prior approaches to ransomware detection or classification. Although malware of a particular kind is called ransomware and many of the previous approaches include ransomware families in the malware dataset, our investigation mainly focuses on the binary and multiclass classification of ransomware through the dynamic analysis approach. First, we present recent research on API sequence and frequency-based ransomware detection and classification techniques. Next, we introduce a few investigations on network traffic features-based methods. Then, we mention several works that combine other significant features along with API call features and network traffic features towards ransomware detection and classification. All of these approaches are similar to our method since we consider both the API call features and network traffic features for comparing the performance of Machine Learning models with and without the RFECV feature selection technique.

A good number of researchers analyzed API call behaviors and proposed ransomware detection or classification methods based on the API call sequences or frequencies. Maniath et al. [10] analyzed the API call behavior of 157 Ransomware and presented LSTM-based ransomware detection that focuses on API call sequence and compensates for the ransomware that causes execution delays. However, this work lacks complete information about the ransomware families/variants and the number of benign software used for the experiment. Vinayakumar Kumar et al. [11] proposed an MLP-based ransomware detection method focusing on API call frequency but they deployed a simple MLP network that failed to distinguish CryptoWall and Cryptolocker.  Z. Chen et al. [27] used API Call Flow Graph (CFG) generated from the extracted API sequence using the API monitor tool for detecting ransomware. Regardless, the work is based on a smaller dataset that includes only four ransomware families. Also, graph-similarity analysis requires higher computational power that some systems may not provide. Takeuchi et al. [12] used API call sequences to identify zero-day ransomware attacks and the work involved kernel tricks for tuning Support Vector Machine. However, the accuracy of this work decreases while using standardized vector representation because of the less diverse dataset. Bae et al. [28] extracted the API call sequences using the Intel Pin Tool. Their sequential process includes generating an n-gram sequence, input vector, and Class Frequency Non-Class Frequency (CF-NCF) for every sample before fitting their model. Nevertheless, their work lacks complete information about the ransomware families/variants used for the experiment, and the work’s accuracy can be improved with the help of deception-based techniques. Hwang et al. [13] analyzed API calls and used two Markov chains, one for ransomware and another for benign software to capture the API call sequence patterns. By using Random Forest, they compensate Markov Chains and control FPR and FNR to achieve better performance. However, their model produces high FPR that can be improved with the help of signature-based techniques.

In contrast to the API call behaviors, some researchers analyzed network traffic behaviors of different ransomware families. Cabaj et al. [29] proposed two real-time Software Defined Networking (SDN) based mitigation methods that were developed using OpenFlow to ensure the prompt reaction to the threat while not decreasing the overall network performance. However, the proposed method is only based on the features of CryptoWall ransomware. Tseng et al. [30] proposed a method that can identify specific network traffic types and detect in-network behavior sequences. Their approach detects ransomware before encryption starts. Regardless, the work lacks complete information about the ransomware families/variants as well as benign software used for the experiment. Alhawi et al. [31] used TShark for capturing and analyzing malicious network traffic activities followed by utilizing the WEKA ML tool to detect ransomware based on only 9 extracted features. Nonetheless, because of using fewer features of only 210 ransomware, the proposed method may fall short of recognizing the new ransomware variants. Almashhadani et al. [24] built a dedicated testbed for executing and capturing the network traffic of the sample ransomware and proposed a multi-classifier that works on two different levels: packet-based and flow-based classifiers. Their method employed a language-independent algorithm that can detect domain names from general sonic axioms. However, the proposed method is only based on the Locky ransomware.

Instead of considering only API call behavior or only the network traffic behavior, some researchers combined these two categories of behavior along with other malicious indicators (i.e., registry key operations, file extensions, files/directory operation, etc.) for their models. D. Sgandurra et al. [9] analyzed API calls, registry key operations, embedded strings, file extensions, files/directory operations, and dropped file extensions prior to developing their model. The features were selected using the mutual information criterion and their proposed method ‘EldeRan’ was able to deal with sophisticated encryption methods of ransomware at an early stage. However, the limitation of ‘EldeRan’ is that it produces a higher False Positive Rate. Continella et al. [32] analyzed filesystem operations and presented two models: process-centric trained on each process and system-centric trained on the whole system. They developed ‘ShieldFS’-a software on OS that can detect malicious file activities and roll back from the attack. However, their system-centric model produces high false positives, and the system may face performance degradation due to the add-on driver on the OS. T. Lu et al. [33] analyzed API calls, network features, registry operations, file operations, directory operations, and memory usage for developing a ransomware detection method based on the Artificial Immune System (AIS). They applied real-valued detector generation based on the V-detector negative selection while optimizing the AIS parameter (i.e., hypersphere detector distribution) to improve the ransomware detection rate. Regardless, their system also produces higher false alarms. Hasan et al. [1] considered API calls, network features, registry key operations. process operations, function length frequency, and printable string information for their model. They proposed a framework- ‘RansHunt’ that takes a hybrid approach to identify potential static and dynamic features for the SVM classifier that outperforms traditional AV tools. However, the proposed method only focuses on the Crypto category. So, it may not be effective for the Locker category.

Table 1 presents the synopsis of the previous research works conducted on the analysis, detection, and classification of ransomware.

\begin{table}
\caption{Synopsis of the Literature Review.}
\label{table1}
\setlength{\tabcolsep}{4pt}
\begin{tabular}{|p{45pt}|p{75pt}|p{50pt}|p{50pt}|}
\hline
\multicolumn{4}{|p{220pt}|}{API Call Features} \\
\hline
Reference & Dataset & Classifier & Accuracy \%\\
\hline
Maniath et al. [10]&
\begin{itemize}[leftmargin=*, topsep=0pt]
    \item 157 Ransomware
    \item Unspecified number of benign software
\end{itemize}&
Long Short-Term Memory&
96.67 $\%$ \\
\hline
Vinayakumar Kumar et al. [11]&
\begin{itemize}[leftmargin=*, topsep=0pt]
    \item 755 Ransomware
    \item 219 Benign Software
\end{itemize}&
Multilayer Perceptron&
\begin{itemize}[leftmargin=*, topsep=0pt]
    \item 100\% (Binary Classification)
    \item 98\% (Multi-class Classification)
\end{itemize}\\
\hline
Z. Chen et al. [27]&
\begin{itemize}[leftmargin=*, topsep=0pt]
    \item 83 Ransomware
    \item 85 Benign Software
\end{itemize}&
Simple Logistic&
98.2\% \\
\hline
Takeuchi et al. [12]&
\begin{itemize}[leftmargin=*, topsep=0pt]
    \item 276 Ransomware
    \item 312 Benign Software
\end{itemize}&
Support Vector Machine&
97.48\% \\
\hline
Bae et al. [28]&
\begin{itemize}[leftmargin=*, topsep=0pt]
    \item 1000 Ransomware
    \item 900 Malware
    \item 300 Benign software
\end{itemize}&
Random Forest&
98.65\% \\
\hline
Hwang et al. [13]&
\begin{itemize}[leftmargin=*, topsep=0pt]
    \item 1909 Ransomware
    \item 1139 Benign software
\end{itemize}&
Markov Chain, Random Forest (Two-stage detection model)&
97.3\% \\
\hline
\multicolumn{4}{|p{220pt}|}{Network Features} \\
\hline
K. Cabaj et al. [29]&
\begin{itemize}[leftmargin=*, topsep=0pt]
    \item 359 CryptoWall samples
\end{itemize}&
N/A&
N/A \\
\hline
Tseng et al. [30]&
\begin{itemize}[leftmargin=*, topsep=0pt]
    \item 155 Ransomware
    \item Unspecified number of benign software
\end{itemize}&
Deep Neural Network&
93.92\% \\
\hline
Alhawi et al. [31]&
\begin{itemize}[leftmargin=*, topsep=0pt]
    \item 210 Ransomware
    \item 264 Benign software
\end{itemize}&
J48&
97.1\% \\
\hline
Almashhadani et al. [24]&
\begin{itemize}[leftmargin=*, topsep=0pt]
    \item Locky ransomware
    \item Unspecified number of benign software
\end{itemize}&
Bayes Net&
99.83\% \\
\hline
\multicolumn{4}{|p{220pt}|}{API Call Features, Network Features, and Other Features} \\
\hline
D. Sgandurra et al. [9]&
\begin{itemize}[leftmargin=*, topsep=0pt]
    \item 582 Ransomware
    \item 942 Benign Software
\end{itemize}&
Regularized Logistic Regression&
96.34\% \\
\hline
A. Continella et al. [32]&
\begin{itemize}[leftmargin=*, topsep=0pt]
    \item 383 Ransomware
    \item 2245 Benign Software
\end{itemize}&
Random Forest&
97.70\% \\
\hline
T. Lu et al. [33]&
\begin{itemize}[leftmargin=*, topsep=0pt]
    \item 1000 Ransomware
    \item 1000 Benign Software
\end{itemize}&
V-detector&
90\% \\
\hline
Hasan et al. [1]&
\begin{itemize}[leftmargin=*, topsep=0pt]
    \item 360 Ransomware
    \item 532 Malware
    \item 460 Benign Software
\end{itemize}&
Support Vector Machine&
97.10\% \\
\hline
\end{tabular}
\label{tab1}
\end{table}

\section{Methodology}
The methodology of our study consists of three subsequent steps as illustrated in Figure 2: Data Collection, Feature Engineering, and Classification.

\begin{figure}[htb]
	\makebox[\linewidth][c]{\includegraphics[angle = 0, clip, trim=0cm 0cm 0cm 0cm, width=0.5\textwidth]{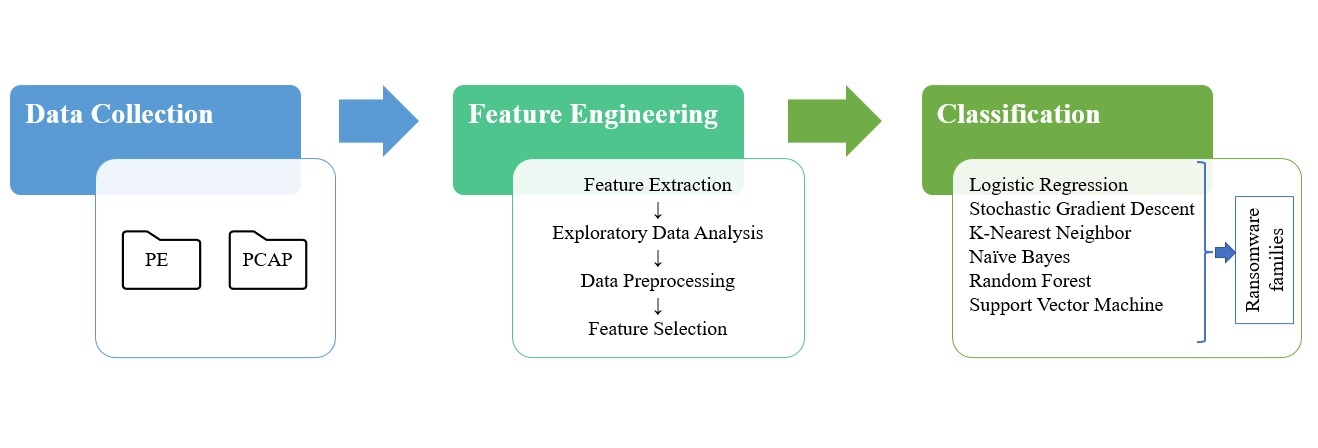}}
        \caption{Process overview of our methodology.}
	\label{fig:pic2}
	\vspace{-10pt}
\end{figure}

\subsection{Data Collection}
We have developed a Web-Crawler- ‘GetRansomware’ to automate collecting the Windows Portable Executable (PE) files of 15 different ransomware families from the VirusShare repository. [34]. We have also shared the Web-Crawler on our GitHub repository for public access [35]. About 95\% of the PE files were collected from VirusShare using GetRansomware. The rest of the PE files were collected from theZoo [36] and Hybrid-Analysis.com [37]. In addition, we have collected the Packet Capture (PCAP) files of those ransomware families from the malware-traffic-analysis [38]. Every ransomware sample was downloaded as a password-protected compressed file. Table 2 presents the number of collected samples.

\begin{table}
\centering
\caption{Number of collected samples.}
\label{table2}
\setlength{\tabcolsep}{4pt}
\begin{tabular}{|p{60pt}|p{30pt}|p{40pt}|}
\hline
\textbf{Ransomware}&
\textbf{PE file}& \textbf{PCAP file} \\
\hline
Cerber (c0) &	95 & 58 \\
\hline
Eris (c1) &	95 & 55 \\
\hline
CryptoWall (c2) &	97 & 55 \\
\hline
Eris (c3) &	98 & 55 \\
\hline
Hive (c4) &	100 & 56 \\
\hline
Jigsaw (c5) &	95 & 60 \\
\hline
Locky (c6) &	95 & 60 \\
\hline
Maze (c7) &	100 & 55 \\
\hline
Mole (c8) &	100 & 56 \\
\hline
Sage (c9) &	100 & 56 \\
\hline
Satan (c10) &	100 & 60 \\
\hline
Shade (c11) &	98 & 57 \\
\hline
TeslaCrypt (c12) &	97 & 59 \\
\hline
Virlock (c13) &	95 & 57 \\
\hline
WannaCry (c14) & 95 & 57 \\
\hline
Total &	1460 & 856 \\
\hline
\end{tabular}
\label{tab2}
\vspace{-10pt}
\end{table}

\subsection{Feature Engineering}
The scarcity of the ransomware dataset is one of the major challenges that hinder the research work in this area [39]. Therefore, for this study, we construct two different datasets from two types of binaries through separate feature engineering processes. In the first process, we create the first dataset by analyzing the PE files while in the second process, we create the second dataset by analyzing the PCAP files.

\subsubsection{Process 1: Creation of the first dataset- ‘Data1’}
The feature engineering step for the first process is composed of two phases. The phases are:

\begin{itemize}
    \item Phase 1: Feature Extraction 
    \item Phase 2: Feature Selection
\end{itemize}

\paragraph{Phase 1: Feature Extraction}
From the wide range of distinct behavioral features, we have considered utilizing Application Programming Interface (API) call frequencies for our study. API calls are made by the application or program running at a user level to request services as depicted in Figure 3. It is the method through which data or information is exchanged between the sending device and the receiving device. The OS performs the requested services by issuing these calls, and the outcomes are returned to the caller user applications. Thus, API calls made by the ransomware program allow the attackers to explore and obtain control of the system and perform malicious activities. Since analyzing API call behavior leads researchers to better understand the program’s behavior [40], [41], therefore, we have considered extracting the API call frequency by executing the PE files of the ransomware.

\begin{figure}[htb]
	\makebox[\linewidth][c]{\includegraphics[angle = 0, clip, trim=0cm 0cm 0cm 0cm, width=0.5\textwidth]{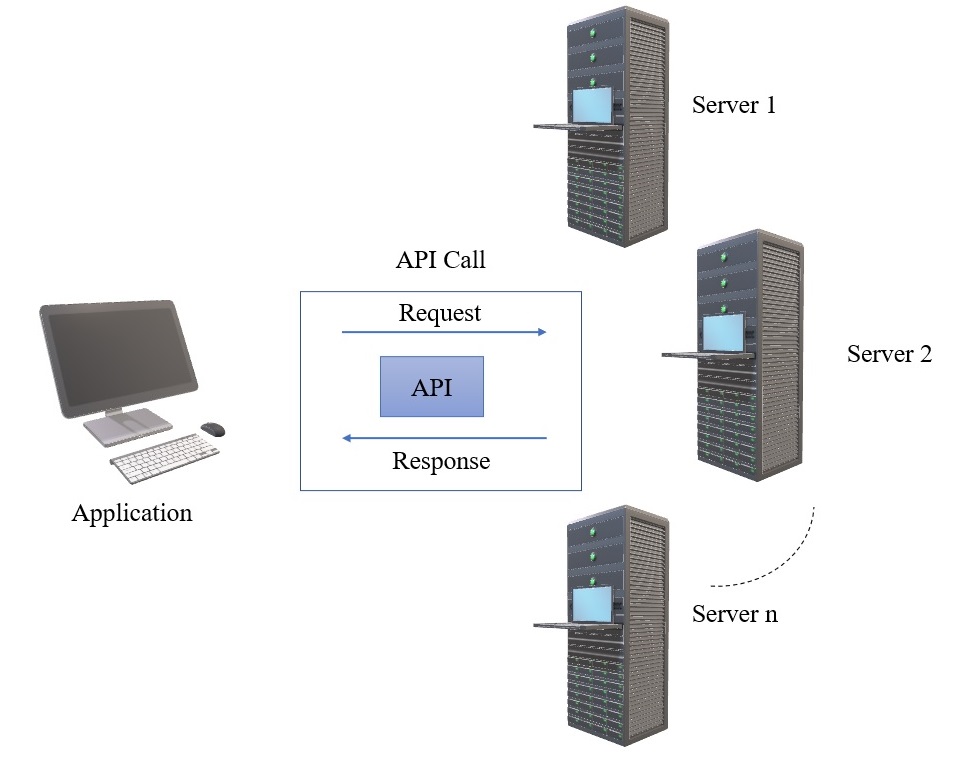}}
        \caption{Communication through the API call.}
	\label{fig:pic3}
	\vspace{-10pt}
\end{figure}

We have analyzed the PE files with the help of Hybrid-Analysis.com [37], powered by the CrowdStrike Falcon Sandbox [42]. To automate submitting malicious binaries, pull the analysis report after the analysis, and perform advanced or required search queries on the database, Falcon Sandbox provides a free, convenient, and efficient API key that one can obtain from an authorized user account. For analysis, we have used our API key and Falcon Sandbox Python API Connector- VxAPI wrapper [43] to automatically submit the binaries from the system. After submission, Falcon Sandbox runs the binaries in a Virtual Machine (VM) and captures the run-time behaviors as illustrated in Figure 4. Later, it shows the analysis results on the web interface.

\begin{figure}[htb]
	\makebox[\linewidth][c]{\includegraphics[angle = 0, clip, trim=0cm 0cm 0cm 0cm, width=0.5\textwidth]{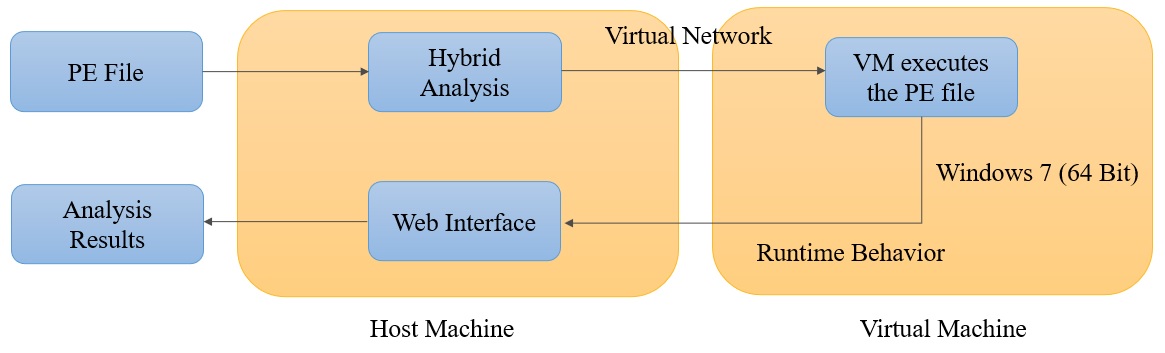}}
        \caption{Block diagram of the PE file execution process.}
	\label{fig:pic4}
	\vspace{-10pt}
\end{figure}

Contrary to the prior works where the analysis tasks were done using the Cuckoo Sandbox [1], [9] - [13], we have analyzed the PE files using the Falcon Sandbox that uses a VM (Windows 7 64-bit) to execute the PE files. Falcon Sandbox incorporates many other services, such as VirusTotal, Thug honeyclient, OPSWAT Metadefender, TOR, NSRL (Whitelist), Phantom, and a large number of antivirus engines to provide an integrated and in-depth analysis reports compared to other Sandboxes. While executing the binaries, we have set run-time to the maximum available duration in the Falcon Sandbox to deal with the delayed execution techniques deployed by the attackers. The total time for the analysis was (1460 PE files * 7 minutes) = 170 hours = 7 days approximately. Next, we obtained the analysis report by using the API key from which we have only sorted and computed the frequency of each API call. At the end of the PE files analysis process, we obtained our first dataset- ‘Data1’ consisting of the different frequencies of 68 distinct API calls associated with the 15 ransomware families as presented in Table 3.

\begin{table}
\caption{List of features in the 'Data1' dataset.}
\label{table3}
\setlength{\tabcolsep}{4pt}
\begin{tabular}{|p{105pt}|p{115pt}|}
\hline
\multicolumn{2}{|p{220pt}|}{API Call Features} \\
\hline
1. FindWindowExW & 35. NtProtectVirtualMemory \\
2. LdrGetDllHandle & 36. NtQueryAttributesFile \\
3. NtAdjustPrivilegesToken & 37. NtQueryDefaultLocale \\
4. NtAlertThread & 38. NtQueryDirectoryFile \\
5. NtAllocateVirtualMemory & 39. NtQueryInformationFile \\
6. NtAlpcSendWaitReceivePort & 40. NtQueryInformationProcess \\
7. NtConnectPort & 41. NtQueryInformationToken \\
8. NtCreateEvent & 42. NtQueryKey \\
9. NtCreateFile & 43. NtQueryObject \\
10. NtCreateKey & 44. NtQuerySystemInformation \\
11. NtCreateKeyEx & 45. NtQueryValueKey \\
12. NtCreateMutant & 46. NtQueryVirtualMemory \\
13. NtCreateSection & 47. NtQueryVolumeInformationFile \\
14. NtCreateThreadEx & 48. NtReadFile \\
15. NtCreateUserProcess & 49. NtReadVirtualMemory \\
16. NtDelayExecution & 50. NtRequestWaitReplyPort \\
17. NtDeleteValueKey & 51. NtResumeThread \\
18. NtDeviceIoControlFile & 52. NtSetContextThread \\
19. NtEnumerateKey & 53. NtSetInformationFile \\
20. NtEnumerateValueKey & 54. NtSetInformationKey \\
21. NtFsControlFile & 55. NtSetInformationProcess \\
22. NtGetContextThread & 56. NtSetInformationThread \\
23. NtMapViewOfSection & 57. NtSetSecurityObject \\
24. NtNotifyChangeKey & 58. NtSetValueKey \\
25. NtOpenDirectoryObject & 59. NtTerminateProcess \\
26. NtOpenEvent & 60. NtTerminateThread \\
27. NtOpenFile & 61. NtUnmapViewOfSection \\
28. NtOpenKey & 62. NtWaitForMultipleObjects \\
29. NtOpenKeyEx & 63. NtWriteFile \\
30. NtOpenMutant & 64. NtWriteVirtualMemory \\
31. NtOpenProcess & 65. NtYieldExecution \\
32. NtOpenProcessToken & 66. OpenSCManager \\
33. NtOpenSection & 67. OpenServiceW \\
34. NtOpenThreadToken & 68. SetWindowsHookEx \\
\hline
\end{tabular}
\label{tab1}
\end{table}


\paragraph{Phase 2: Feature Selection}
At the beginning of the feature selection phase, we have evenly divided (stratified train-test split) our dataset into train data (80\%) and test data (20\%) to avoid data leakage. Next, we have applied Recursive Feature Elimination with Cross-Validation (RFECV) [48] to our train data. RFECV is a wrapper-style feature selection method that wraps a given ML model as depicted in Figure 5 and selects the optimal number of features for each model by recursively eliminating 0-n features in each loop. Next, it selects the best-performing subset of features based on the accuracy or the score of cross-validation. RFECV also removes the dependencies and collinearity existing in the model. By using RFECV, we have selected 6 distinct subsets of features for 6 ML classifiers. These features have been selected by setting ‘min features to select’ as 34 (half of the features), cv=5, and ‘scoring’= ‘accuracy’ so that RFECV would select at least half of the features based on the optimum accuracy over the 5-fold cross-validation.

\begin{figure}[htb]
	\makebox[\linewidth][c]{\includegraphics[angle = 0, clip, trim=0cm 0cm 0cm 0cm, width=0.5\textwidth]{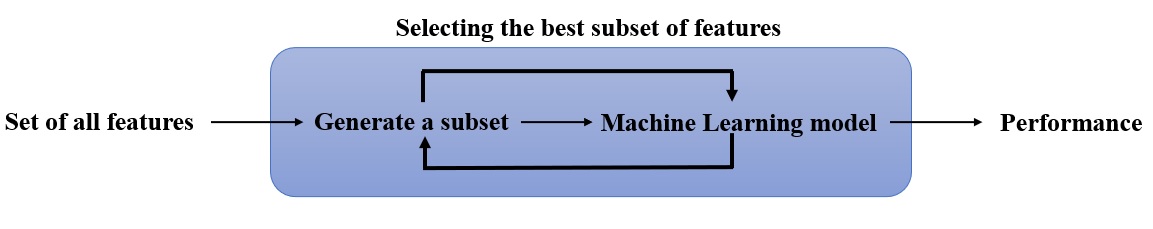}}
        \caption{RFECV feature selection technique.}
	\label{fig:pic5}
	\vspace{-10pt}
\end{figure}

\subsubsection{Process 2: Creation of the second dataset- ‘Data2’}
The feature engineering step for the second process is composed of four phases. The phases are:

\begin{itemize}
    \item Phase 1: Feature Extraction
    \item Phase 2: Exploratory Data Analysis (EDA)
    \item Phase 3: Data Preprocessing
    \item Phase 4: Feature Selection
\end{itemize}

\paragraph{Phase 1: Feature Extraction}
We have considered utilizing network traffic features for the second dataset for our study. The Transmission Control Protocol (TCP) refers to the set of standardized communication protocols that specify how computers communicate over the network. According to our literature review, the communication between the infected host machine (source) and the attacker (destination) is conducted through the transport layer [25]. Besides, HTTP GET or POST methods are also used to send back the information to the attacker [24]. Hence, we have opted for capturing the TCP traffic and the HTTP traffic information by analyzing the PCAP files of the ransomware. 

Again, ransomware often spreads through spam emails containing malignant attachments as macro-enabled word documents. By executing a script, these attachments download the executable file of that ransomware from a URL and install it on the system. After the installation, the ransomware continuously tries to search and connect to its C\&C servers to exchange the encryption key and launch the attack session. Firstly, it utilizes an encrypted list of IP addresses for creating a TCP session with the C\&C servers. Upon failure due to the unreachable or blacklisted IP addresses or disrupted session, the ransomware then opts to find out its C\&C server by executing the Domain Generation Algorithm (DGA) and recurrently produces a good number of pseudo-random domain names. Then, the ransomware continues sending the Domain Name System (DNS) request to those domain names until the actual C\&C server is found as illustrated in Figure 6. Here, DNS converts human-readable domain names to machine-readable IP addresses. Upon successful establishment of a TCP session, the attacker guides the victim in delivering the payload. The characteristic of dispatching an extensive number of DNS requests looking for a real C\&C server looks like an arbitrary set of characters. Meaningful statistical information can be derived from these requested domain names as well as the pattern of randomness found in them [22]. If the ransomware detection method can trace the randomness that occurs before finding out the actual C\&C server, it can be stopped before the ransomware begins encrypting files. This is an efficient approach in case of a zero-day attack as deriving the information from the known ransomware is not required in this case. Therefore, we have opted for extracting DNS traffic information by analyzing the PCAP files of the ransomware.

\begin{figure}[htb]
	\makebox[\linewidth][c]{\includegraphics[angle = 0, clip, trim=0cm 0cm 0cm 0cm, width=0.5\textwidth]{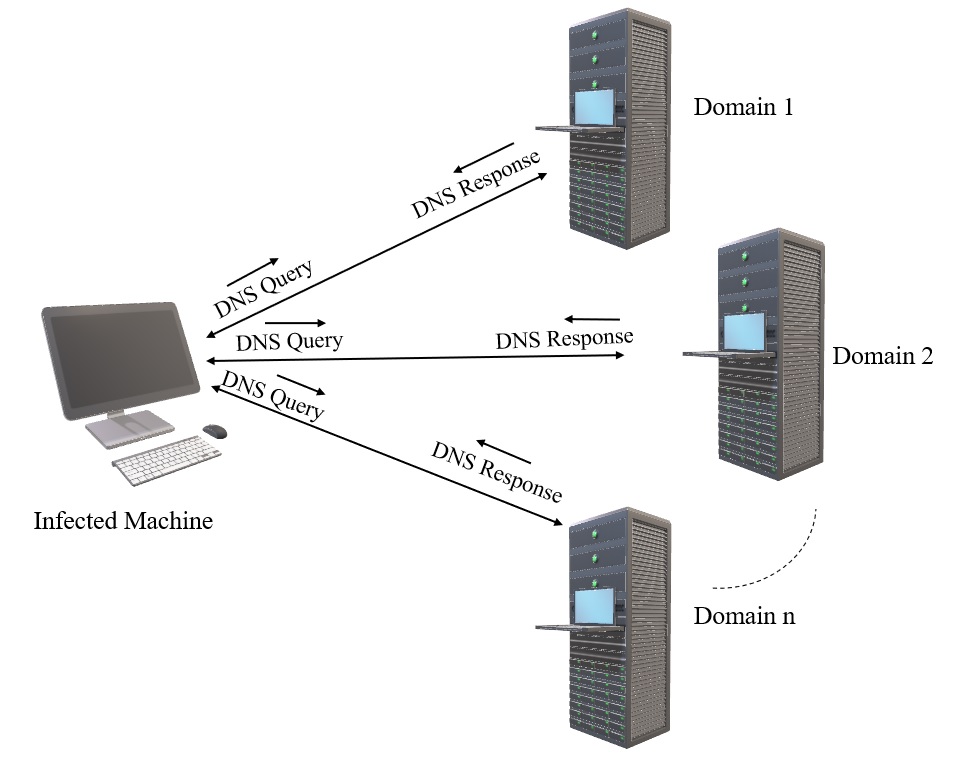}}
        \caption{Finding out the actual C\&C server by sending DNS requests.}
	\label{fig:pic6}
	\vspace{-10pt}
\end{figure}

We have analyzed the PCAP files using Wireshark- a network protocol analyzer [44], [45]. This manual process involved three identical systems with Wireshark installed and 2 volunteers for analyzing the PCAP files. We have extracted 18 network traffic features that according to [46], convey important statistical information that enhances the ability of the classification algorithms to classify ransomware. Then, these features have been merged resulting in ‘Data2’. Table 4 presents the list of network traffic features.

\begin{table}[htb]
\centering
\caption{List of features in the 'Data2' dataset.}
\label{table4}
\setlength{\tabcolsep}{4pt}
\begin{tabular}{|p{180pt}|}
\hline
\textbf{Network Traffic Features} \\
\hline
1. IP and port of the client \\
2. IP and port of the server \\
3. Bytes sent from the client to the server \\
4. Bytes sent from the server to the client \\
5. RSTs in the TCP connection from client to server \\
6. RSTs in the TCP connection from server to client \\
7. FINs in the TCP connection from client to server \\
8. FINs in the TCP connection from server to client \\
9. Number of HTTP requests present in the connection \\
10. HTTP method (GET or POST) of the HTTP requests \\
11. Response code to the HTTP requests \\
12. URL requested in the HTTP request \\
13. Timestamp of the DNS request \\
14. IP and port of the client in the DNS request \\
15. IP and port of the DNS server \\
16. RCode of the DNS response (It is sent by the server indicating whether it was able to settle the request or not) \\
17. DNS request \\
18. DNS response \\
\hline
\end{tabular}
\label{tab2}
\vspace{-10pt}
\end{table}

\paragraph{Phase 2: Exploratory Data Analysis (EDA)}
At the beginning of Phase 2, we have evenly divided (stratified train-test split) the dataset into train data (80\%) and test data (20\%) to avoid data leakage. Next, we have done exploratory data analysis to better understand the raw data so that the data could be preprocessed as per requirement. The findings from this phase are:

\begin{itemize}
    \item Categorical data: We have found 11 features containing categorical data. They are the IP and port of the client, IP and port of the server, Bytes sent from the client to the server, Bytes sent from the server to the client, HTTP method GET or POST of the HTTP requests, Response code to the HTTP requests, URL requested in the HTTP request, IP and port of the client.1, IP and port of the DNS server, DNS request, and DNS response. These categorical data need to be encoded into numerical values since the classifiers require the data to be understandable so that they can be trained on and make predictions.
    \item Random missing values: Since different ransomware families create different numbers of conversations over the network, the number of instances captured from the PCAP files was different for each ransomware sample. Hence, we have observed missing values in network traffic information. Handling missing values is an essential part of the feature engineering process as the ML models may generate biased or inaccurate results if the missing values are not handled properly. There are two ways of dealing with missing values, such as deleting the missing values and imputing the missing values. Since deleting the missing values ends up deleting the entire row or column that contains the missing values, there is a probability of losing useful information in the dataset. So, we have opted for imputing the missing values.
\end{itemize}

\paragraph{Phase 3: Data Preprocessing}
In the data preprocessing phase, firstly, we have encoded the categorical data into numerical data for which we have applied One-Hot Encoding [47] by using the ‘.get\_dummies’ attribute of Pandas data frame package that generates the dummy variables of those 11 features. For preventing the ‘Dummy Variable Trap’, we have set ‘True’ as ‘drop\_first’ parameter. To normalize the data and to prevent the imputer from producing biased numerical replacements for the missing data, we have scaled the numerical values between 0 and 1. After normalizing the data, we have used Scikit-Learn’s Impute package to apply KNNImputer to fill up the missing values.

\paragraph{Phase 4: Feature Selection}
We have selected the network traffic features using RFECV by setting ‘min\_features\_to\_select’ as 9 (half of the features), cv=5, and ‘scoring’= ‘accuracy’ so that RFECV would select at least half of the features based on the optimum accuracy over the 5-fold cross-validation applied on our train data.

\subsection{Classification}
We have employed Supervised Machine Learning algorithms to classify 15 ransomware families into corresponding categories. Supervised learning algorithms are trained on the labeled dataset to make a decision in response to the unseen test dataset. These algorithms are generally of two types, such as classification-based and regression-based. The classification-based algorithms are used to accomplish both binary and multi-class classification where the instances from the test dataset are classified into one among an array of known classes, such as Naïve Bayes, Random Forest, K-Nearest Neighbor, etc. On the other hand, regression-based algorithms consider the relationship between independent features or input variables and dependent target class or continuous output variables to make a prediction, such as Linear Regression, Neural Network Regression, Lasso Regression, etc. As this study focuses on classifying 15 ransomware families, the following algorithms have been employed that are widely used for both binary and multi-class classification as per requirement:

\begin{itemize}
    \item Logistic Regression (LR): is a type of statistical analysis that predicts the probability of a dependent variable from a set of independent variables using their linear combination.
    \item Stochastic Gradient Descent (SGD): is an optimization algorithm to find the model parameters by updating them for each training data so that the best fit is reached between predicted and actual outputs.
    \item K-Nearest Neighbor (KNN): estimates the likelihood of a new data point being a member of a specific group by measuring the distance between neighboring data points and the new data point.
    \item Naïve Bayes (NB): is based on Bayes’ theorem and predicts the probability of an instance belonging to a particular class.
    \item Random Forest (RF): constructs multiple decision trees during the training phase and finally determines the class selected by the maximum number of trees.
    \item Support Vector Machine (SVM): takes one or more data points from different classes as inputs and generates hyperplanes as outputs that best distinguish the classes.
\end{itemize}

Since this study focuses on multi-class classification and some classifiers are only designed for binary classification problems (i.e., Logistic Regression, Support Vector Machine, etc.), these cannot be directly applied to multi-class classification problems. Therefore, Heuristic Methods [52] can be applied to divide a multi-class classification problem into several binary classification problems. There are two types of heuristic methods as illustrated in Figure 7. The methods are:

\begin{itemize}
    \item One-vs-Rest (OvR) which splits the dataset into one class against all other classes each time [53].
    \item One-vs-One (OvO) which splits the dataset into one class against every other class each time [54].
\end{itemize}

We have applied the OvR method for our experiment to reduce the time and computational complexities. All these classifiers are built along with ‘RandomSearchCV’ [51]- a hyperparameter optimization technique, to find the best combination of hyperparameters for maximizing the performance of the models’ output in a reasonable time. Instead of exhaustively searching for the optimal values of the hyperparameters through a manually determined set of values (i.e., Grid Search), RandomSearchCV randomly searches the grid space and selects the best combination of hyperparameter values based on the accuracy or the score of cross-validation. Since we have used RFECV for feature selection and RandomSearchCV for hyperparameter optimization, the Nested Cross-Validation technique has been implemented in the pipeline to build each model.

\begin{figure}[htb]
	\makebox[\linewidth][c]{\includegraphics[angle = 0, clip, trim=0cm 0cm 0cm 0cm, width=0.5\textwidth]{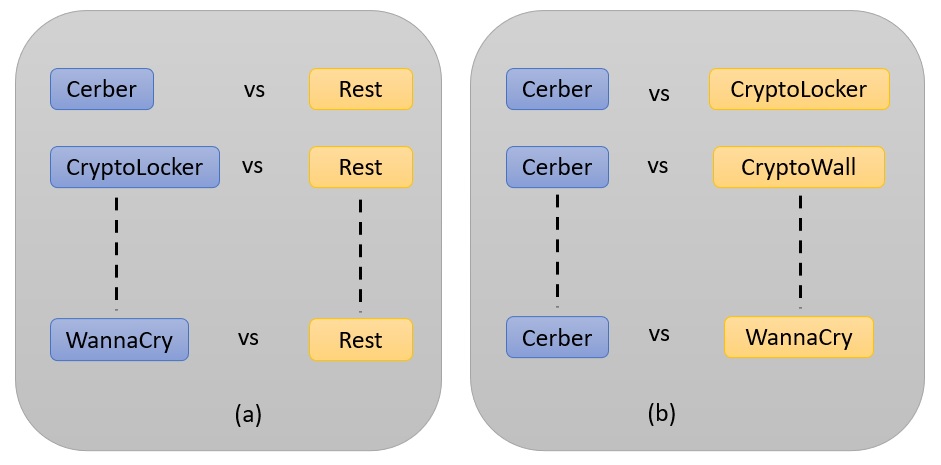}}
        \caption{Heuristic methods: (a) One-vs-Rest and (b) One-vs-One.}
	\label{fig:pic7}
	\vspace{-10pt}
\end{figure}

\section{Experimental Results and Discussions}
\subsection{Experimental Results}
We have evaluated the models in terms of Precision, Recall, F1-score, and Accuracy. These performance metrics are measured as follows:

\begin{equation*}
\begin{aligned}
& Precision = \frac{TP}{TP+FP} \\
& Recall = \frac{TP}{TP + FN} \\
& F1-score = \frac{2 \times Precision \times Recall}{Precision + Recall} \\
& Accuracy = \frac{TP+TN}{TP + TN +FP + FN} \times 100\\
\label{eq1}
\end{aligned}
\end{equation*}
\noindent where, TP = True Positives, FP = False Positives (Type 1 Error), TN = True Negative, FN = False Negative (Type 2 Error).

\begin{table*}
\caption{Performance comparison between LR, SGD, KNN, NB, RF, and SVM with respect to the with-feature selection and without-feature selection using the ‘Data1’ dataset (W FS= With-Feature Selection, and W/O FS= Without-Feature Selection).}
\label{table5}
\begin{tabular}{|p{1.6cm}|p{0.7cm}|p{1cm}|p{0.7cm}|p{1cm}|p{0.7cm}|p{1cm}|p{0.7cm}|p{1cm}|p{0.7cm}|p{1cm}|p{0.7cm}|p{1cm}|}
\hline
Performance & \multicolumn{2}{c|}{LR} & \multicolumn{2}{c|}{SGD} & \multicolumn{2}{c|}{KNN} & \multicolumn{2}{c|}{NB} & \multicolumn{2}{c|}{RF} & \multicolumn{2}{c|}{SVM} \\
\cline{2-13}
 & W FS &	W/O FS &	W FS &	W/O FS &	W FS &	W/O FS &	W FS &	W/O FS &	W FS &	W/O FS &	W FS &	W/O FS\\
\hline
$Accuracy_{avg}$ &	98.20 &	99.30 &	90.43 &	92.45 &	89.62 &	90.52 &	97.17 &	97.46 &	91.51 &	92.78 &	94.34 &	95.58\\
\hline
$Precision_{avg}$ &	98.53 &	99.37 &	98.86 &	100 &	94.33 & 	94.08 &	97.82 &	98.77 &	100 &	99.79 &	99.15 &	99.21\\
\hline
$Recall_{avg}$ &	98.22 &	99.30 &	91.36 &	92.45 &	89.62 &	90.52 &	97.17 &	97.46 &	91.51 &	92.78 &	94.34 &	95.58\\
\hline
$F1{\text -}score_{avg}$ &	98.20 &	99.29 &	94.61 &	95.85 &	90.78 &	91.27 &	97.21 &	98.02 &	95.23 &	95.87 &	96.40 &	97.19\\

\hline
\end{tabular}
\end{table*}

Table 5 presents the performance comparison of Machine Learning models with and without feature selection for the ‘Data1’ dataset. It shows that with and without feature selection LR outperforms other classifiers securing 98.20\% and 99.30\% overall accuracy respectively. Although there is a slight performance degradation in all the classifiers in the with-feature selection scenario, remarkable improvement in the processing time has been observed. As shown in Table 6, with-feature selection, the average processing time of all the classifiers has been improved by 26.97\%. We present the classification accuracy for each class of the best-performed supervised machine learning model from these classifiers in two different scenarios. Figure 8 illustrates the normalized confusion matrix of the LR classifier. As shown in Figure 8(a), when the features are not selected, among 15 classes, the classifier could distinguish 13 classes with 100\% accuracy. However, the classifier produces 1\% false negatives classifying CryptoLocker ransomware and 11\% false positives classifying Shade ransomware. On the other hand, Figure 8(b) shows the confusion matrix of the LR classifier with feature selection. Although the classifier could distinguish 10 classes with 100\% accuracy, the classifier produces 1\% false negatives classifying Cerber, 22\% false positives classifying CryptoLocker, 10\% false positives classifying Mole, 10\% false positives classifying Sage, and 11\% false positives classifying Shade ransomware.

\begin{table}
\caption{Classifier’s processing time comparison without-feature selection and with-feature selection using the ‘Data1’ dataset.}
\label{table6}
\setlength{\tabcolsep}{4pt}
\begin{tabular}{|p{50pt}|p{50pt}|p{50pt}|p{50pt}|}
\hline
Classifier &	Without-feature selection (in seconds) & 	With-feature selection (in seconds) & 	Improvement (\%)\\
\hline
LR &	79.21 &	58.44 &	26.22\\
\hline
SGD &	78.43 &	57.62 &	26.53\\
\hline
KNN &	78.00 &	51.25 &	34.29\\
\hline
NB &	76.39 &	55.67 &	27.12\\
\hline
RF &	75.41 &	58.28 &	22.71\\
\hline
SVM &	79.19 &	59.43 &	24.95\\
\hline
\multicolumn{3}{|c|}{Average processing time improvement (\%)} & 26.97\\
\hline
\end{tabular}
\label{tab1}
\end{table}

\Figure[t!](topskip=0pt, botskip=0pt, midskip=0pt)[width=0.85\textwidth]{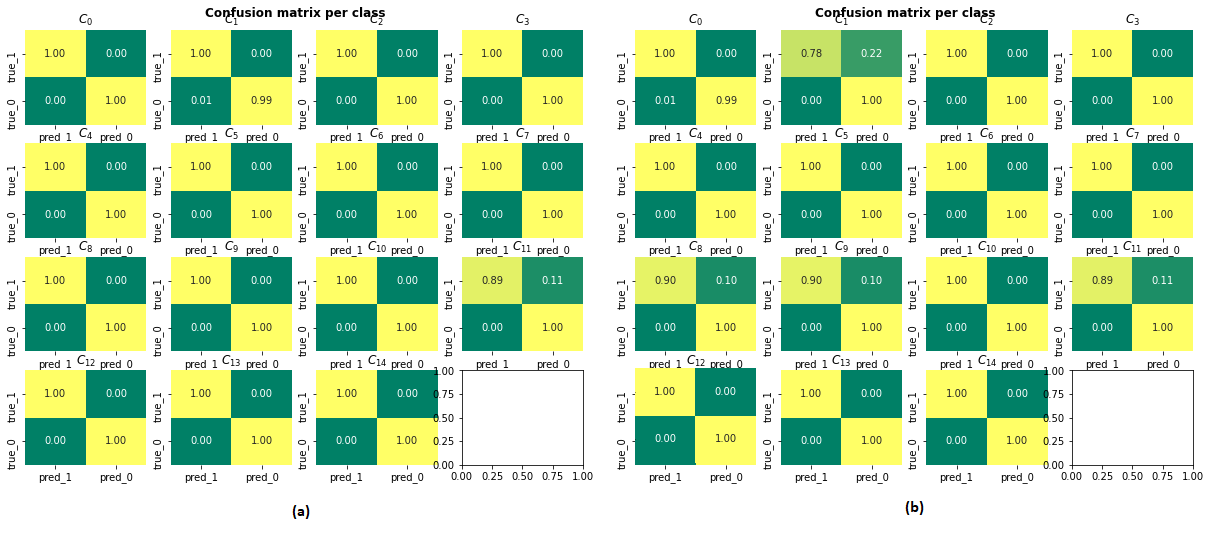}
{Confusion matrix of (a) Logistic Regression without feature selection, and (b) Logistic Regression with feature selection for the ‘Data1’ dataset.\label{fig:pic8}}

Table 7 presents the performance comparison of Machine Learning models with and without feature selection for the ‘Data2’ dataset. It shows that with and without feature selection NB outperforms other classifiers securing 97.89\% and 98.95\% overall accuracy respectively. Even though all of the classifiers in the with-feature selection scenario show a minor performance deterioration, a notable improvement in processing time has been seen. As shown in Table 8, with-feature selection, the average processing time of all the classifiers has been improved by 34.72\%. We present the classification accuracy for each class of the best-performed supervised machine learning model from these classifiers in two different scenarios. Figure 9 illustrates the normalized confusion matrix of the NB classifier. As shown in Figure 9(a), when the features are not selected, among 15 classes, the classifier could distinguish 10 classes with 100\% accuracy. However, the classifier produces 2\% false negatives classifying CryptoLocker and 1\% false negatives classifying Maze ransomware.  On the other hand, Figure 9(b) shows the confusion matrix of the NB classifier with feature selection. The classifier could distinguish 9 classes with 100\% accuracy with no false negatives. However, with feature selection, the classifier produces higher false positives as compared to that without-feature selection.

\begin{table*}
\caption{Performance comparison between LR, SGD, KNN, NB, RF, and SVM with respect to the with-feature selection and without-feature selection using the ‘Data2’ dataset (W FS= With-Feature Selection, and W/O FS= Without-Feature Selection).}
\label{table7}
\begin{tabular}{|p{1.6cm}|p{0.7cm}|p{1cm}|p{0.7cm}|p{1cm}|p{0.7cm}|p{1cm}|p{0.7cm}|p{1cm}|p{0.7cm}|p{1cm}|p{0.7cm}|p{1cm}|}
\hline
Performance & \multicolumn{2}{c|}{LR} & \multicolumn{2}{c|}{SGD} & \multicolumn{2}{c|}{KNN} & \multicolumn{2}{c|}{NB} & \multicolumn{2}{c|}{RF} & \multicolumn{2}{c|}{SVM} \\
\cline{2-13}
 & W FS &	W/O FS &	W FS &	W/O FS &	W FS &	W/O FS &	W FS &	W/O FS &	W FS &	W/O FS &	W FS &	W/O FS\\
\hline
$Accuracy_{avg}$ & 92.25 &	94.04 &	81.69 &	82.76 & 80.99 &	83.25 &	97.89 &	98.95 &	78.87 &	79.96 &	92.25 &	93.90\\
\hline
$Precision_{avg}$ & 98.21 &	97.81 &	90.96 & 93.27 & 92.05 & 92.56 &	98.21 &	99.05 &	100 &	99.90 &	98.73 &	98.89\\
\hline
$Recall_{avg}$ &	92.25 &	94.04 &	88.03 &	87.53 & 80.99 &	83.25 &	97.89 &	98.95 &	78.87 &	79.96 &	92.25 &	93.90\\
\hline
$F1{\text -}score_{avg}$ & 94.81 &	95.67 &	88.98 & 89.91 &	84.13 &	85.96 &	97.92 &	98.95 &	86.99 &	87.64 &	95.01 &	96.06\\
\hline
\end{tabular}
\end{table*}

\begin{table}
\caption{Classifier’s processing time comparison without-feature selection and with-feature selection using the ‘Data2’ dataset.}
\label{table8}
\setlength{\tabcolsep}{4pt}
\begin{tabular}{|p{50pt}|p{50pt}|p{50pt}|p{50pt}|}
\hline
Classifier &	Without-feature selection (in seconds) & 	With-feature selection (in seconds) & 	Improvement (\%)\\
\hline
LR &	88.19 &	56.88 &	35.5\\
\hline
SGD &	85.31 &	54.66 &	35.9\\
\hline
KNN &	85.44 &	54.30 &	36.4\\
\hline
NB &	84.13 &	51.29 &	35.5\\
\hline
RF &	85.27 &	56.78 &	33.4\\
\hline
SVM &	83.18 &	56.93 &	31.6\\
\hline
\multicolumn{3}{|c|}{Average processing time improvement (\%)} & 34.72\\
\hline
\end{tabular}
\label{tab1}
\end{table}

\Figure[t!](topskip=0pt, botskip=0pt, midskip=0pt)[width=0.85\textwidth]{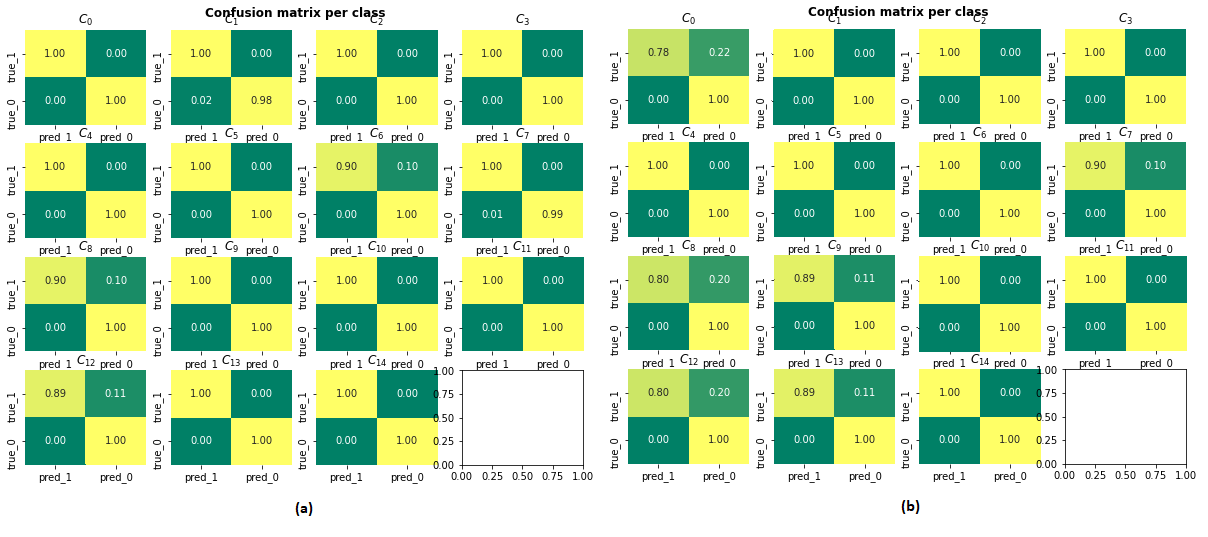}
{Confusion matrix of (a) Naïve Bayes without feature selection, and (b) Naïve Bayes with feature selection for the ‘Data2’ dataset.\label{fig:pic9}}

\subsection{Discussions}
In this section, we present the comparison between the RFECV-selected features in the with-feature selection scenario and the highly contributing features in the without-feature selection scenario to examine the efficiency of the RFECV feature selection technique toward ransomware classification. For this task, we apply ‘Shapley Additive exPlanations’, a tool for visualizing data that helps explain the results of machine learning models. SHAP is based on the coalition game theory that measures each feature’s individual contribution to the final output while conserving the sum of contributions being the same as the final result [26]. When it comes to the performance evaluation of any model, knowing both ‘What’ and ‘Why’ the models have taken these decisions is equally important. The answer to ‘What’ presents the results or outputs of the machine learning models while the answer to ‘Why’ explains the factors, or features affecting the results. While some predictive models may not require explainability because of their usage in a low-risk real-world environment, some models that deal with the real-world high-risk environment (i.e., ransomware detection/classification) need explanation. Unlike other explanation techniques that are limited to explaining specific models, SHAP values can be used to explain a wide variety of models, such as DeepExplainer to explain Deep Neural Networks (i.e., Multi-Layer Perceptron, Convolutional Neural Networks, etc.), TreeExplainer to explain tree-based models (i.e., Random Forest, XGBoost, etc.), and KernelExplainer to explain any model, etc. [49], [50]. For our study, we have used TreeExplainer to obtain highly contributing features from the Random Forest classifier, while for the other classifiers we have used KernelExplainer. 

For the classification model, the SHAP value is regarded as a 2-D array where the columns represent the features used in the model and the rows represent individual predictions predicted by the model. So, each SHAP value in this array indicates a specific feature’s contribution to that row’s prediction output, as shown in Figure 10. Here, a positive SHAP value specifies that a feature is positively pushing the base value or expected value to the model output. On the other hand, the negative SHAP value specifies that the feature is negatively pushing the base value to the model output. The base value or the mean model output is computed based on the train data.

\begin{figure}[htb]
	\makebox[\linewidth][c]{\includegraphics[angle = 0, clip, trim=0cm 0cm 0cm 0cm, width=0.45\textwidth]{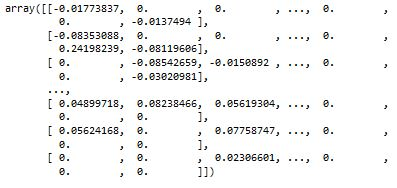}}
        \caption{Array of SHAP values.}
	\label{fig:pic10}
	\vspace{-10pt}
\end{figure}

Passing the array of SHAP values to a ‘summary plot’ function creates a feature importance plot as shown in Figure 11. Here, we illustrate 40 highly contributing features (as RFECV selects the highest 40 features for the KNN classifier) for each classifier in the without-feature selection scenario for the ‘Data1’ dataset. Here, the x-axis denotes the mean of the absolute SHAP value for each feature which indicates the total contribution of the feature to the model and the y-axis denotes the features used for the classification. The features are organized in descending order from top to bottom by how strongly they influence the model’s decision. As illustrated in Figure 11, the set of highly contributing features and their order varies for each classifier. However, for our study, we only examine the variation of the RFECV-selected features with the highly contributing features of the corresponding classifiers. Table 9 presents the set of optimum features selected by RFECV for each ML classifier from the ‘Data1’ dataset and Table 10 presents the list of RFECV-selected features for each ML classifier that is not present in the top 40 highly contributing features. By comparing these two tables, we get the features that are causing performance deterioration in the with-feature selection scenario and producing higher false alarms as compared to that without-feature selection. 

\Figure[t!](topskip=0pt, botskip=0pt, midskip=0pt)[width=1.0\textwidth]{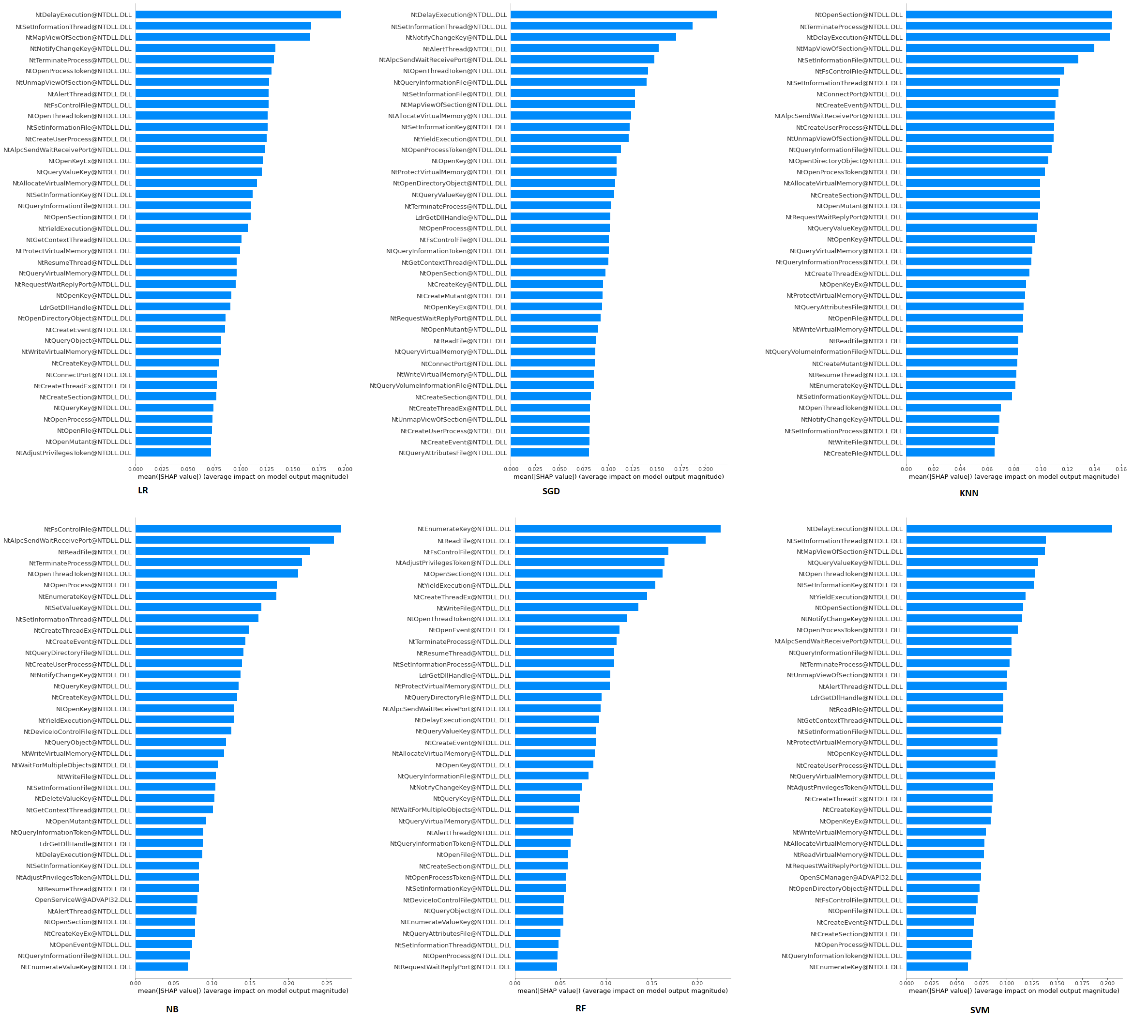}
{Summary plot showing the top 40 highly contributing features of the ‘Data1’ dataset for each ML classifier in the without feature selection scenario.\label{fig:pic11}}

\DefTblrTemplate{contfoot-text}{normal}{}
\SetTblrTemplate{contfoot-text}{normal}

\begin{longtblr}[
  caption = {Set of optimum features selected by RFECV from the ‘Data1’ dataset for each ML classifier.},
  label = {table9},
]{
  colspec = {|XX|},
  rowhead = 1,
  cell{1,2}{1} = {c=2}{l}, 
  hspan = minimal,
}

\hline
\textbf{Selected API Call features}\\
\hline
\textbf{Classifier:} &	LR\\
\textbf{Total Selected Features:} & 38 \\
\hline
NtDelayExecution & NtSetinformationThread \\
NtMapViewOfSection & NtNotifyChangeKey\\
NtTerminateProcess & NtOpenProcessToken\\
NtUnmapViewOfSection & NtAlertThread\\
NtFsControlFile & NtOpenThreadToken\\
NtSetinformationFile & NtCreateUserProcess\\
NtAIpcSendWaitReceivePort & NtOpenKeyEx\\
NtAllocateVirtualMemory & NtSetInformationKey\\
NtQueryInformationFile & NtYieldExecution\\
NtProtectVirtualMemory & NtResumeThread\\
NtQueryVirtualMemory & NtRequestWaitReplyPort\\
LdrGetDIlHandle & NtEnumerateKey\\
NtOpenEvent & NtQueryInformationProcess\\
NtOpenDirectoryObject & NtCreateEvent\\
NtWriteVirtualMemory & NtCreateKey\\
NtConnectPort & NtCreateThreadEx\\
NtCreateSection & NtQuerykey\\
NtOpenProcess & NtOpenFile\\
NtOpenMutant & NtAdjustPrivilegesToken\\
\hline
\textbf{Classifier:} &	SGD\\
\textbf{Total Selected Features:} & 38 \\
\hline
NtDelayExecution &	NtSetinformationThread\\
NtNotifyChangeKey & NtAlertThread\\
NtAIpcSendWaitReceivePort & NtOpenThreadToken\\
NtQueryInformationFile & NtSetInformationFile\\
NtMapViewOfSection & NtAllocateVirtualMemory\\
NtSetinformationKey & NtYieldExecution\\
NtOpenProcessToken & NtOpenkey\\
NtProtectVirtualMemory & NtQueryValueKey\\
NtTerminateProcess & NtOpenProcess\\
NtFsControlFile & NtQueryInformationToken\\
NtOpenSection & NtCreateKey\\
NtCreateMutant & NtRequestWaitReplyPort\\
NtOpenMutant & NtReadFile\\
NtConnectPort & NtWriteVirtualMemory\\
NtCreateSection & NtCreateThreadEx\\
NtUnmapViewOfSection & NtCreateEvent\\
NtQueryAttributesFile & NtEnumerateKey\\
NtOpenEvent & NtQueryInformationProcess\\
NtSetContextThread & NtQueryDirectoryFile\\
\hline
\textbf{Classifier:} &	KNN\\
\textbf{Total Selected Features:} & 40 \\
\hline
NtOpenSection & NtTerminateProcess\\
NtDelayExecution & NtMapViewOfSection\\
NtSetinformationFile & NtFsControlFile\\
NtSetinformationThread & NtCreateEvent\\
NtAIpcSendWaitReceivePort & NtCreateUserProcess\\
NtUnmapViewOfSection & NtQueryInformationFile\\
NtOpenDirectoryObject & NtAllocateVirtualMemory\\
NtCreateSection & NtOpenMutant\\
NtQueryValueKey & NtOpenKey\\
NtQueryInformationProcess & NtCreateThreadEx\\
NtProtectVirtualMemory & NtQueryAttributesFile\\
NtOpenFile & NtWriteVirtualMemory\\
NtReadFile & NtQueryVolume-lnformationFile\\
NtCreateMutant & NtResumeThread\\
NtEnumerateKey & NtSetInformationKey\\
NtOpenThreadToken & NtNotifyChangeKey\\
NtSetinformationProcess & NtWriteFile\\
NtCreateFile & NtDeviceIoControlFile\\
NtAlertThread & NtGetContextThread\\
NtQueryDirectoryFile & NtSetContextThread\\
\hline
\textbf{Classifier:} &	NB\\
\textbf{Total Selected Features:} & 38 \\
\hline
NtAllocateVirtualMemory  &  NtFsControlFile\\
NtAlpcSendWaitReceivePort & NtReadFile\\
NtTerminateProcess & NtOpenThreadToken\\
NtOpenProcess & NtEnumerateKey\\
NtSetValueKey & NtSetinformationThread\\
NtCreateThreadEx & NtCreateEvent\\
NtQueryDirectoryFile & NtCreateUserProcess\\
NtNotifyChangeKey & NtQueryKey\\
NtCreateKey & NtOpenkey\\
NtYieldExecution & NtDeviceloControlFile\\
NtQueryObject & NtWriteVirtualMemory\\
NtWaitForMultipleObjects & NtDeleteValueKey\\
NtGetContextThread & NtOpenMutant\\
NtQueryInformationToken & NtDelayExecution\\
NtSetInformationKey & NtAdjustPrivilegesToken\\
NtResumeThread & OpenServiceW\\
NtAlertThread & NtOpenSection\\
NtCreateKeyEx & NtOpenEvent\\
NtQueryInformationFile & NtEnumerateValueKey\\
\hline
\textbf{Classifier:} &	RF\\
\textbf{Total Selected Features:} & 38 \\
\hline
NtEnumerateKey & NtReadFile\\
NtFsControlFile & NAdjustPrivilegesToken\\
NtOpenSection & NtYieldExecution\\
NtCreateThreadEx & NtWriteFile\\
NtOpenThreadToken & NtOpenEvent\\
NtTerminateProcess & NtResumeThread\\
NtSetinformationProcess & NtProtectVirtualMemory\\
NtQuery DirectoryFile & NtAIpcSendWaitReceive-Port\\
NtDelayExecution & NtCreateEvent\\
NtAllocateVirtualMemory & NtQueryInformationFile\\
NtQueryKey & NtWaitForMultipleObjects\\
NtQueryVirtualMemory & NtAlertThread\\
NtOpenFile & NtOpenProcess Token\\
NtSetinformationKey & NtDeviceloControlFile\\
NtQueryobject & NtEnumerateValueKey\\
NtQueryAttributesFile & NtSetinformationThread\\
NtOpenProcess & NtRequestWaitReplyPort\\
NtCreateUserProcess & NtGetContextThread\\
NtQueryInformationProcess & NtMapViewOfSection\\
\hline
\textbf{Classifier:} &	SVM\\
\textbf{Total Selected Features:} & 37 \\
\hline
NtDelayExecution &  NtSetinformationThread\\
NtMapViewOfSection & NtQueryValueKey\\
NtOpenThreadToken & NtSetinformationKey\\
NtOpenSection & NtNotifyChangeKey\\
NtOpenProcessToken & NtAIpcSendWaitReceive-Port\\
NtTerminate rocess & NtUnmapViewOfSection\\
NtAlertThread & LdrGetDIlHandle\\
NtGetContextThread & NtSetinformationFile\\
NtOpenKey & NtCreateUserProcess\\
NtQueryVirtualMemory & NtAdjustPrivilegesToken\\
NtCreateThreadEx & NtCreateKey\\
NtWriteVirtualMemory & NtAllocateVirtualMemory\\
NtReadVirtualMemory & NtRequestWaitReplyPort\\
OpenSCManager & NtOpenDirectoryObject\\
NtFsControlFile & NtOpenFile\\
NtCreateEvent & NtCreateSection\\
NtOpenProcess & NtQueryInformationToken\\
NtEnumerateKey & NtSetContextThread\\
NtDeviceIoControlFile & \\
\hline
\end{longtblr}

\begin{table}
\caption{List of RFECV-selected features from the ‘Data1’ dataset for each ML classifier that is not present in the top 40 highly contributing features.}
\label{table10}
\setlength{\tabcolsep}{4pt}
\begin{tabular}{|p{30pt}|p{90pt}|p{30pt}|p{50pt}|}
\hline
Classifier &	API Call Features &	Total & 	Average Performance Decrease with RFECV-selected features (\%)\\
\hline
LR &	NtEnumerateKey  &	3 &	1.10\\
& NtOpenEvent & &\\
& NtQueryInformationProcess & &\\
\hline
SGD &	NtEnumerateKey &	5 &	2.02\\
& NtOpenEvent & &\\
& NtQueryInformationProcess & &\\
& NtSetContextThread & &\\
& NtQueryDirectoryFile & &\\
\hline
KNN &	NtDeviceIoControlFile &	5 &	0.9\\
& NtAlertThread & &\\
& NtGetContextThread & &\\
& NtQueryDirectoryFile & &\\
& NtSetContextThread & &\\
\hline
NB &	NtAllocateVirtualMemory &	1 &	0.29\\
\hline
RF &	NtCreateUserProcess &	4 &	1.27\\
& NtGetContextThread & &\\
& NtQueryInformationProcess & &\\
& NtMapViewOfSection & &\\
\hline
SVM &	NtSetContextThread &	2 &	1.24\\
& NtDeviceIoControlFile & &\\
\hline
\end{tabular}
\label{tab1}
\end{table}

Similarly, for the ‘Data2’ dataset, we present the comparison between the RFECV-selected features in the with-feature selection scenario and the highly contributing features in the without-feature selection scenario. Figure 12 illustrates the features of the ‘Data2’ dataset in descending order from top to bottom by how strongly they influence the model’s decision. For each classifier, the order of the features varies except for the ‘Bytes sent from the client to the server’ feature. However, similar to the previous step, we only examine the variation of the RFECV-selected features. Table 11 presents the set of optimum features selected by RFECV from the ‘Data2’ dataset for each ML classifier, and Table 12 presents the list of features that were not selected by the RFECV. By comparing these two tables, we get the features that are causing performance deterioration even with the best-performed ML classifier in the with-feature selection scenario and produce higher false alarms as compared to that without-feature selection.

Although SHAP importance shows the effect of a given feature on the model output while disregarding the exactness of the prediction, our study, by comparing the highly contributing features in the without feature selection scenario and the RFECV selected features in the with feature selection scenario finds out that the RFECV feature selection technique, sometimes fails to select the features that have a high impact on the model output resulting in performance degradation. Again, for two different ransomware datasets, the selected features have been ranked 1, while the not-selected features have been ranked greater than 1. Therefore, the order of the selected features based on their importance remains unknown in the RFECV feature selection technique making it less efficient in ransomware classification. 

\Figure[t!](topskip=0pt, botskip=0pt, midskip=0pt)[width=1.0\textwidth]{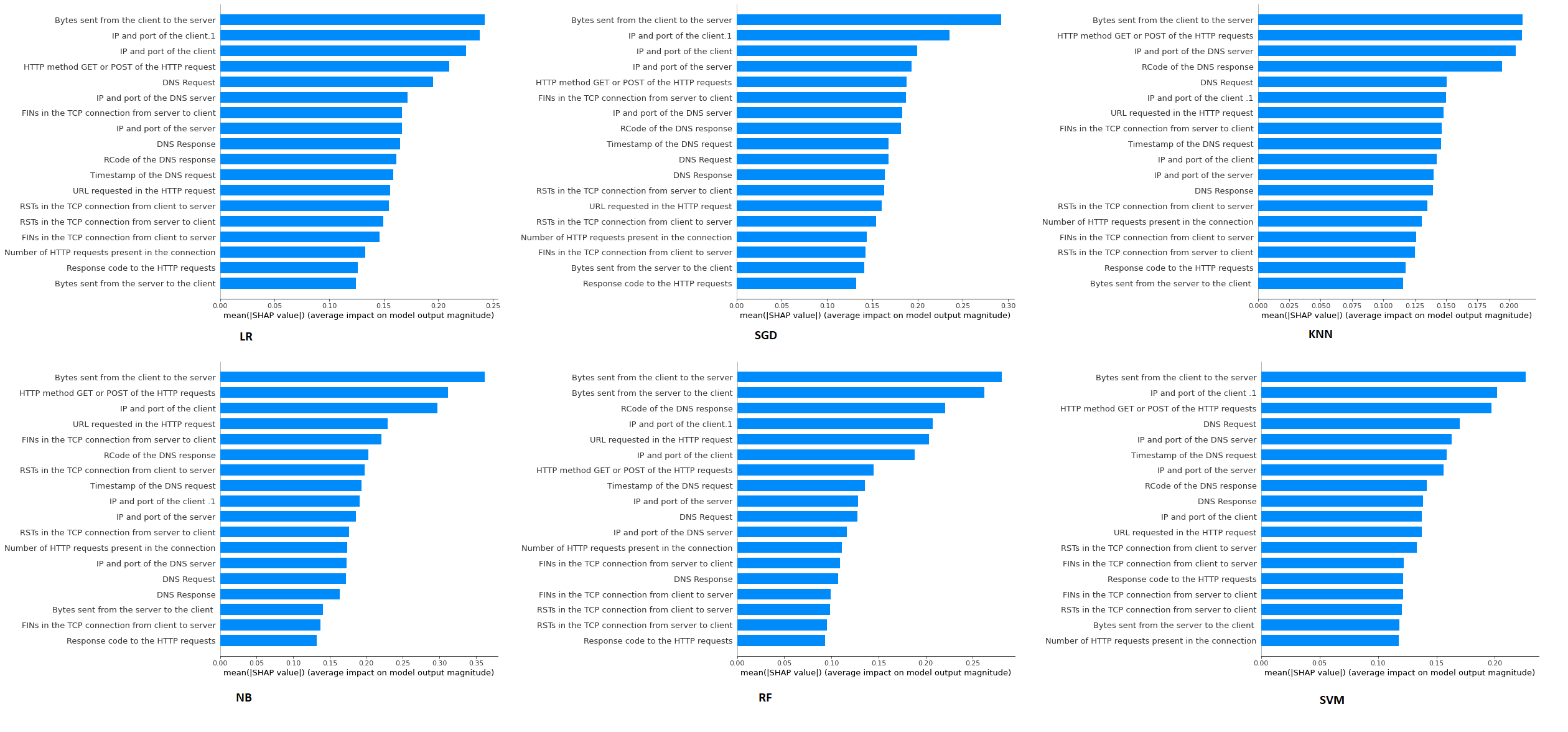}
{Summary plot showing the features of the ‘Data2’ dataset in descending order based on their contribution to each ML classifier’s decision. \label{pic12.png}}

\begin{longtblr}[
  caption = {Set of the optimum number of features selected by RFECV from the ‘Data2’ dataset for each ML classifier.},
  label = {table11},
]{
  colspec = {|X|},
  rowhead = 1,
}

\hline
\textbf{Selected Network Traffic Features}\\
\hline
\textbf{Classifier:} LR\\
\textbf{Number of Selected Network Traffic Features:} 14\\
\hline
RSTs in the TCP connection from server to client\\
RSTs in the TCP connection from client to server\\
FINs in the TCP connection from client to server\\
FINs in the TCP connection from server to client\\
Bytes sent from the client to the server\\
Bytes sent from the server to the client\\
DNS Request\\
DNS Response\\
IP and port of the client.1\\
IP and port of the client\\
HTTP method GET or POST of the HTTP request\\
IP and port of the server\\
Timestamp of the DNS request\\
RCode of the DNS response\\
\hline
\textbf{Classifier:} SGD\\
\textbf{Number of Selected Network Traffic Features:} 10\\
\hline
IP and port of the client.1\\
IP and port of the client\\
IP and port of the server\\
FINs in the TCP connection from server to client\\
URL requested in the HTTP request\\
RSTs in the TCP connection from client to server\\
Bytes sent from the client to the server\\
HTTP method GET or POST of the HTTP requests\\
Number of HTTP requests present in the connection\\
Response code to the HTTP requests\\
\hline
\textbf{Classifier:} KNN\\
\textbf{Number of Selected Network Traffic Features:} 10\\
\hline
Bytes sent from the client to the server\\
Bytes sent from the server to the client\\
HTTP method GET or POST of the HTTP requests\\
Response code to the HTTP requests\\
IP and port of the DNS server\\
RCode of the DNS response\\
URL requested in the HTTP request\\
IP and port of the client\\
RSTs in the TCP connection from client to server\\
Number of HTTP requests present in the connection\\
\hline
\textbf{Classifier:} NB\\
\textbf{Number of Selected Network Traffic Features:} 16\\
\hline
RCode of the DNS response\\
RSTs in the TCP connection from client to server\\
IP and port of the server\\
RSTs in the TCP connection from server to client\\
Number of HTTP requests present in the connection\\
IP and port of the DNS serve\\
DNS Request\\
DNS Response\\
Bytes sent from the server to the client\\
FINs in the TCP connection from client to server\\
Response code to the HTTP requests\\
Bytes sent from the client to the server\\
HTTP method GET or POST of the HTTP requests\\
IP and port of the client\\
URL requested in the HTTP request\\
FINs in the TCP connection from server to client\\
\hline
\textbf{Classifier:} RF\\
\textbf{Number of Selected Network Traffic Features:} 13\\
\hline
Timestamp of the DNS request\\
IP and port of the DNS server\\
Number of HTTP requests present in the connection\\
FINs in the TCP connection from server to client\\
Bytes sent from the client to the server\\
Bytes sent from the server to the client\\
RCode of the DNS response\\
IP and port of the client.1\\
URL requested in the HTTP request\\
IP and port of the client\\
HTTP method GET or POST of the HTTP requests\\
DNS Response\\
RSTs in the TCP connection from server to client\\
\hline
\textbf{Classifier:} SVM\\
\textbf{Number of Selected Network Traffic Features:} 13\\
\hline
RSTs in the TCP connection from client to server\\
RSTs in the TCP connection from server to client\\
Bytes sent from the client to the server\\
Bytes sent from the server to the client\\
Number of HTTP requests present in the connection\\
IP and port of the client .1\\
HTTP method GET or POST of the HTTP requests\\
DNS Request\\
IP and port of the DNS server\\
Timestamp of the DNS request\\
IP and port of the server\\
RCode of the DNS response\\
DNS Response\\
\hline
\end{longtblr}

\begin{longtblr}[
  caption = {List of features from the ‘Data2’ dataset that were not selected by the RFECV.},
  label = {table12},
]{
  colspec = {|X|X[1.8]|X[0.8]|X[1.4]|},
  rowhead = 1,
}

\hline
Classifier &	Not Selected Network Traffic Features &	Total & 	Average Performance Decrease with RFECV-selected features (\%) \\
\hline
LR  & IP and port of the DNS server & 4 & 1.79\\
& URL requested in the HTTP request & &\\
& Number of HTTP requests present in the connection & &\\
& Response code to the HTTP requests & &\\
\hline
SGD & Timestamp of the DNS request & 8 & 1.07\\
& DNS Request & &\\
& DNS Response & &\\
& RSTs in the TCP connection from server to client & &\\
& IP and port of the DNS server & &\\
& RCode of the DNS response & &\\
& FINs in the TCP connection from client to server & &\\
& Bytes sent from the server to the client & &\\
\hline
KNN & FINs in the TCP connection from client to server & 8 & 2.26\\
& RSTs in the TCP connection from server to client & &\\
& IP and port of the server & &\\
& DNS Response & &\\
& FINs in the TCP connection from server to client & &\\
& Timestamp of the DNS request & &\\
& DNS Request & &\\
& IP and port of the client .1 & &\\
\hline
NB & Timestamp of the DNS request & 2 & 1.06\\
& IP and port of the client .1 & &\\
\hline
RF & FINs in the TCP connection from client to server & 5 & 1.09\\
& RSTs in the TCP connection from client to server & &\\
& IP and port of the server & &\\
& DNS Request & &\\
& Response code to the HTTP requests & &\\
\hline
SVM & FINs in the TCP connection from client to server & 5 & 1.65\\
& Response code to the HTTP requests & &\\
& FINs in the TCP connection from server to client & &\\
& IP and port of the client & &\\
& URL requested in the HTTP request & &\\
\hline
\end{longtblr}

\section{Conclusion}
In this paper, we present a comprehensive performance analysis of widely utilized Supervised Machine Learning models with and without RFECV to quantify the efficiency of this feature selection technique in ransomware classification. Our study finds out that although the classification accuracies are nearly similar in both scenarios, with RFECV the classifiers produce higher false alarms as compared to those without feature selection. In addition, the selected features have been ranked 1 for two separate ransomware datasets, whereas the not-selected features have been ranked higher. As a result, the RFECV feature selection approach does not reveal the importance-based order in which the features have been chosen. To summarize, by presenting this comparative study, this paper can provide future direction to the researchers in this domain who are looking for efficient feature selection techniques. 





\begin{IEEEbiography}[{\includegraphics[width=1in,height=1.25in,clip,keepaspectratio]{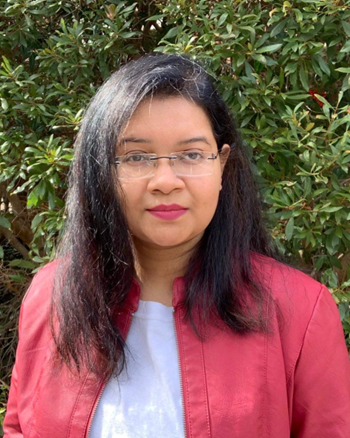}}]{Rawshan Ara Mowri} was born in Dhaka, Bangladesh. She received her Bachelor of Science in Computer Science and Engineering from Bangladesh.  After graduation, she worked as a Math Educator at the British Standard School in Dhaka, Bangladesh. Later, she started pursuing her Master of Science in Computer Science at North Carolina Agricultural and Technical State University in January 2021. During her graduate study, in the first year, she worked as a Graduate Research Assistant at the Human-Computer Interactions and Biometrics Lab. The following year, she joined the Center for Cyber Defense (CCD) Lab as a Graduate Research Assistant. Her research work is heavily focused on applying human-understandable machine learning in cyber security, data privacy, and big data analytics.
\end{IEEEbiography}

\begin{IEEEbiography}[{\includegraphics[width=1in,height=1.25in,clip,keepaspectratio]{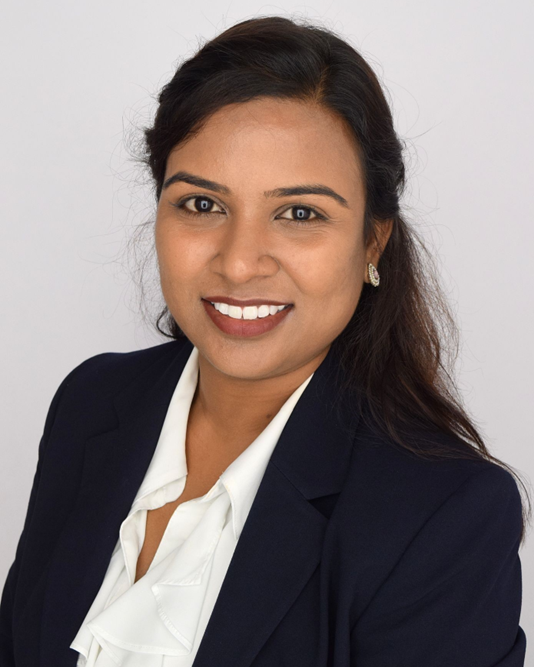}}]{Madhuri Siddula} received a B.S. degree in computer science engineering from Osmania University, India, and an M.S. degree in information security from the Indraprastha Institute of Information Technology, India. She received her Ph.D. degree from the Department of Computer Science, at Georgia State University in 2020. She is currently an Assistant Professor in the Department of Computer Science at North Carolina A\&T State University. Her research interests include social networks, privacy and security, and IoT.
\end{IEEEbiography}

\begin{IEEEbiography}[{\includegraphics[width=1in,height=1.25in,clip,keepaspectratio]{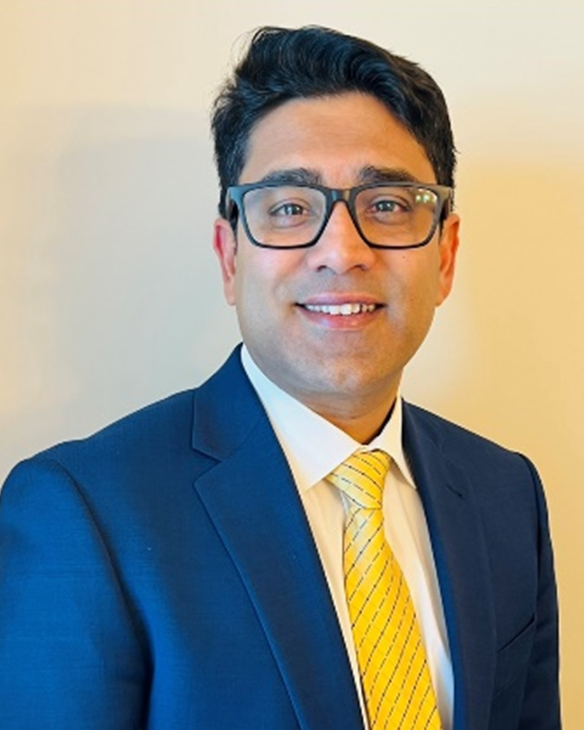}}]{Dr. Kaushik Roy} is a Professor and Interim Chair in the Department of Computer Science at North Carolina A\&T State University (NCAT). His current research is heavily focused on cybersecurity, cyber identity, biometrics, machine learning (deep learning), data science, cyber-physical systems, and big data analytics. He has over 140 publications including 36 journal articles and a book. Dr. Roy has been the PI on \$13.5M and Co-PI on \$9M in research grants funded by the National Science Foundation (NSF), Department of Defense (DoD), National Security Agency (NSA), and Department of Energy (DoE). Dr. Roy is a director of the Center for Cyber Defense (CCD). Dr. Roy also directs the Cyber Defense and AI lab..
\end{IEEEbiography}

\EOD

\end{document}